# Experimental verification of the anomalous skin effect in copper using emissivity measurements

T. Echániz,[1] I. Setién-Fernández,[1] R. B. Pérez-Sáez,[1,2,a)] and M. J. Tello[1,2,3]

[1]*Departamento de Física de la Materia Condensada, Facultad de Ciencia y Tecnología, UPV/EHU, Apdo, 644. 48080 Bilbao, Spain*
[2]*Instituto de Síntesis y Estudio de Materiales, Universidad del País Vasco, Apdo, 644. 48080 Bilbao, Spain*
[3]*CIC Energigune, Parque Tecnológico, Albert Einstein 48, 01510 Miñano, Álava, Spain*

Spectral directional emissivity has been measured in copper between 3 and 24 $\mu$m above room temperature. The experimental spectrum shows a weak broad peak between 7 and 14 $\mu$m, which is much more acute for higher emission angles. However, the peak width and position are both independent of the emission angle. The experimental results are in very good agreement with the semiclassical theory of the optical properties of metals in the regime of the anomalous skin effect, in particular with the asymptotic approximation. This comparison suggests that this work shows an optical experimental evidence of the anomalous skin effect.

The spatial decay of the electromagnetic transverse wave due to the wave-particle interaction is usually referred to as the anomalous skin effect. It is a fundamental phenomenon that is important for a number of applications: lasers, low temperature plasmas, metals, etc.[1–5] For metals, the classical electron theory proposed by Drude[6] gives the reflectance as a function of only two material parameters: the dc-conductivity ($\sigma_0$) and the relaxation time ($\tau$). This approximation is valid when the electron mean free path ($\ell$) is small compared with the classical skin depth ($\delta$). Otherwise the velocity of electrons is proportional to a variable electric field which acts on electrons along its mean free path before reaching $r$. In this case, the system is in the anomalous skin effect regime. Therefore, a simple calculation of $\ell$ and $\delta$ allows to select those metals which are candidates for showing the skin anomalous effect. For example, in the case of copper[7] for $\lambda = 10\,\mu$m, $T = 27\,°$C and $\omega\tau = 5.1$, the following results are obtained: $\delta = 1.2 \times 10^{-6}$ cm and $\ell = 4.2 \times 10^{-6}$ cm. Similar values were found for the other noble metals. These results suggest that, even at room temperature, copper is in the anomalous skin effect regime for medium infrared frequencies.

A semiclassical theory to find the surface impedance related to an arbitrary electric field based on the linearized Boltzmann kinetic equation was developed for specular and diffuse electron scattering at the metal surface as well as for normal and oblique incidence.[3,8–11] A quantum-mechanical treatment[13] gives the same results for the surface impedance that the semi-classical ones. From the surface impedance, it is possible to write the dependence of reflectance on temperature, frequency, and other physical constants. The theory predicts that the directional spectral reflectance ($\rho_\lambda(\theta)$) of metals passes through a broad minimum for infrared frequencies where the metal is in the anomalous skin effect regime. For opaque samples $\rho_\lambda(\theta) = (1 - \epsilon_\lambda(\theta))$, where $\epsilon_\lambda(\theta)$ is the directional spectral emissivity. Therefore, in the anomalous skin effect regime the emissivity must show a broad peak for infrared frequencies. The peak frequency increases when the temperature decreases. However, the height of the peak decreases with increasing temperature. On the other hand, the classical electromagnetic theory for smooth metal surfaces predicts that the emissivity increases with the angle of emission.

The emissivity can be calculated[14,15] by using the experimental optical constants ($\bar{n} = n + ik$). However, the measurement of the optical constants offers serious difficulties. We can even say that good experimental results cannot be expected above 10 $\mu$m.[16] This point explains the large scatter in the relatively small number of literature experimental data about the optical constants of copper,[7,17–19] most of them at room temperature and for $\lambda < 8\,\mu$m. This scatter has its origin in instrumental sources,[16] errors in measured quantities, and differences in the samples derived from the use of different methods of preparation, as well as different surface states. In any case, the copper infrared normal emissivity, calculated in this work from the optical constants data in the literature,[7,18] seems to show a smooth increase above 7 $\mu$m at room temperature. Direct emissivity measurements can be an alternative to avoid uncertainties associated with the infrared optical constants data. This means that highly accurate measurements of copper emissivity in the 3 to 24 $\mu$m spectral range, can clarify the existence of an emissivity broad peak around 10 $\mu$m due to the anomalous skin effect, according to the theoretical predictions. The present Letter shows, by using a high accuracy infrared radiometer, the dependence of the emissivity of copper with wavelength, temperature, and emission angle in the medium infrared. The copper normal spectral emissivity has been studied at its melting point,[20–22] above the melting point,[23] at 400 °C,[24,25] and at liquid helium temperature.[26] However, measures covering the medium infrared spectrum show no signs of the presence of the anomalous skin effect.[24–26] In addition, no directional spectral emissivity data in copper has been measured in the past.

In this work, the experimental measurements were carried out using the homemade HAIRL-radiometer,[27] which allows accurate signal detection as well as its fast processing. A diaphragm adjusts the sample area viewed by the detector

---

[a)]raul.perez@ehu.es

and ensures good temperature homogeneity of the sample measurement area. The sample holder allows directional measurements, and the sample chamber assures a controlled atmosphere (vacuum, inert gas, or open atmosphere). The set-up calibration was carried out by using a modified two-temperature method[28] and the emissivity was obtained by using the blacksur method.[29] The combined standard uncertainty of this direct emissivity device was previously obtained from the analysis of all uncertainty sources.[30]

The samples consisted on electrolytic copper films (>36 $\mu$m) on 6 cm diameter iron discs. Chemical analysis and X-ray diffraction showed that copper had purity above 99.9% in all samples. The surface roughness values for sample 2 were $R_a = 1.22\,\mu$m, $R_z = 7.12\,\mu$m, and $R_t = 8.02\,\mu$m. Similar values were found for the other samples. This copper ensures a minimum signal value, which allows for accurate experimental emissivity data. The roughness gives an intermediate situation between the specular surface reflection ($p=1$) and the completely diffused reflection ($p=0$).

All the measurements were carried out in the 3 to 24 $\mu$m wavelength interval following the same experimental procedure. Once each sample was introduced into the sample chamber, moderate vacuum was made following a slightly reducing low-pressure atmosphere ($N_2 + 5\% H_2$) to prevent oxidation of the sample surface. In addition, two previous heating cycles up to 800 °C ensured that possible sample surface stresses were completely removed.[31,32] Finally, the surface samples were analyzed after five heating cycles by means of X-ray diffraction, optical microscopy, and electron microscopy, and no signs of oxidation were found. The experimental results refer to the third, fourth, or fifth heating cycles.

The experimental emissivity values at 350 °C are plotted in Fig. 1. The experimental uncertainty increases with wavelength being around 2% at 4 $\mu$m and achieving a value of 12% at 16 $\mu$m. It is remarkable the broad peak between 8 and 14 $\mu$m, in agreement with the theoretical predictions for the anomalous skin effect, which has not been previously observed. A comparison between the experimental results and the theoretical predictions for normal incidence, together with the optical constants data in the literature, brings out several interesting features. The model assumes[1] that a fraction $p$ ($0 \leq p \leq 1$) of the electrons that make a collision with the metal surface is specularly reflected, while the rest are diffusely scattered. The difference in reflectance between the $p=1$ and $p=0$ curves could be measurable experimentally with a good reflectometer.[33] On the other hand, the electrical field is not, in general, of exponential form. As a consequence, the classical concept of the complex refractive index loses its physical meaning, and the experimental measurements of the optical constants become complicated. The theoretical calculus[3,10] gives the surface impedance ($Z$) as a function of the physical parameters. Using the asymptotic approximation,[3] the emissivity is given by

$$\epsilon = \epsilon_{clas} \frac{4}{\pi} \sqrt{\frac{2\alpha}{3}} \frac{\omega\tau\mathcal{R}(I) - \mathcal{I}(I)}{(1+\omega^2\tau^2)\sqrt{-\omega\tau + \sqrt{1+\omega^2\tau^2}}}. \quad (1)$$

Here, $\epsilon_{clas}$ is the classical emissivity, and $\mathcal{R}(I)$ and $\mathcal{I}(I)$ stand for the real and imaginary parts of the following integral:

$$I = \int_0^\infty \frac{dt}{t^2 + \frac{i\alpha}{(1+i\omega\tau)^3}k(t)}, \quad (2)$$

where $k(t)$ is a function of $t$ (perpendicular distance to the metal surface),[3] and $\alpha$ is a coefficient given by

$$\alpha = \frac{8\pi^2\omega(em\bar{v})^2\ell^3}{c^2h^3} = \frac{3}{2}\left(\frac{\ell}{\delta}\right)^2. \quad (3)$$

Here, $\omega$ is the frequency, $m$ (electron mass) $\approx m^\star$ (electron effective mass), $\ell = \tau\bar{v}$ the mean free length, $\bar{v}$ the mean velocity, $\tau$ the relaxation time, and $\delta$ the classical skin depth. The other physical parameters are

$$n = \frac{8\pi}{3}\left(\frac{m\bar{v}}{h}\right)^3; \quad \sigma = \frac{ne^2\ell}{m\bar{v}}; \quad \delta = \frac{c}{\sqrt{2\pi\omega\sigma}}; \quad (4)$$

with $n$ the number of electrons per unit volume, $\sigma$ the dc conductivity, and $\delta$ the classical skin depth, respectively.

A plot of expression (1) as a function of $\omega\tau$, keeping $\alpha/\omega\tau$ constant, predicts that the emissivity effectively passes through a broad peak in the far and medium infrared for low and room temperatures, respectively. Since $\bar{v}$ is a constant for a given temperature, the value of $\ell$ also determines $\tau$, and expression (1) represents the frequency variation of the emissivity of a metal at a given temperature. The pure metal value of $\alpha/\omega\tau$ is between 1, above room temperature, and $10^6$ for liquid helium temperature. For copper, above room temperature, for a broad peak at $\lambda = 10\,\mu$m, with a relaxation time $\tau = 1.35^{-14}$s, and a dc conductivity $\sigma = 5.76 \times 10^{17}$ esu, we find $\alpha/\omega\tau = 1.18$. With this value, Eq. (1) is plotted in Fig. 2, together with the theoretical classical emissivity. The maximum emissivity value is 0.0212, and the relative height of the peak ($\epsilon_{peak} - \epsilon_{clas}$) is 0.00140 for $\lambda = 10.13\,\mu$m. In Fig. 3, the theoretical and experimental wavelength dependence of ($\epsilon_{peak} - \epsilon_{clas}$) are shown for comparison. It must be noticed that, in this case, the $\epsilon_{clas}$ experimental values are obtained from a simple interpolation of the emissivity values out of the broad peak wavelength range.

The theoretical peak width is larger than the experimental one, whereas the shape of the peak is the same in both cases. The markedly different width of these two peaks is due to the fact that $\epsilon_{clas}$ experimental values are calculated by means of a base line so far as $\lambda = 20\,\mu m$, while the theoretical ones are extended to $100\,\mu m$. Therefore, differences must be expected since an asymptotic approach is being used. In any case, the plots in Fig. 3 present a very high qualitative agreement. A re-examination[10] of the theory of the anomalous skin effect for normal incidence yields a similar peak width but the $(\epsilon_{peak} - \epsilon_{clas})$ values are lower than those in Fig. 3.

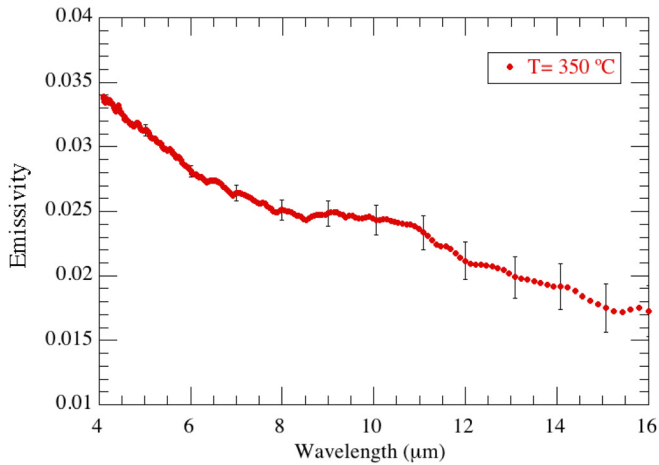

FIG. 1. Measured emissivity as a function of wavelength at 350 °C.

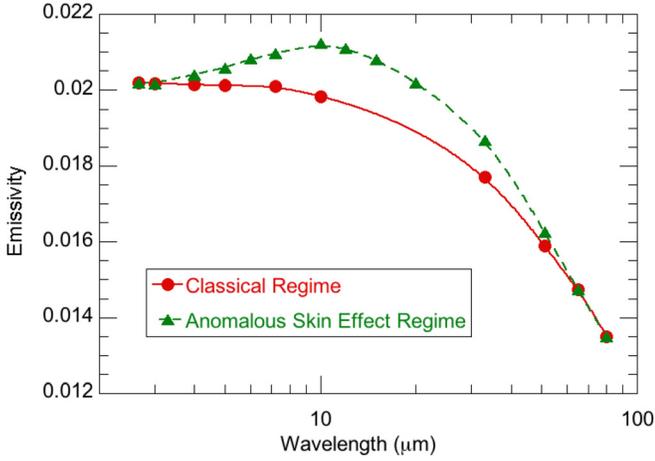

FIG. 2. Theoretical copper emissivity curves above room temperature (250 °C). Broken line shows the theoretical behaviour in the anomalous skin effect regime, and solid line stands for classical skin effect.

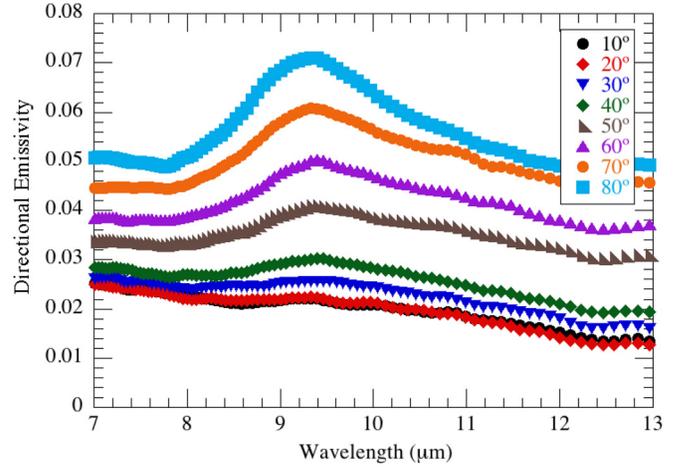

FIG. 4. Spectral emissivity results at 250 °C for eight emission angles. The angular dependence is in agreement with the electromagnetic theory and with the asymptotic approach of the anomalous skin effect theory.

The directional emissivity measurements of copper with average polarization (Fig. 4) show a clear dependence of the broad peak on the incidence angle $\theta$. The theoretical surface impedance predictions[3,10] showed that the angular dependence is nearly negligible in the anomalous skin effect regime. This theoretical result seems to be in contradiction with the experimental plot in Fig. 4. However, the experimental results are in agreement with the theoretical prediction if we take into account that, according to the general electromagnetic theory, the emissivity must show an increase when the angle increases. Therefore, in this case, the peak width must be the same for the $80° \geq \theta \geq 0°$ incidence angle interval, and the value of $(\epsilon_{peak} - \epsilon_{clas})$ for each frequency of the broad peak must show nearly the same increase with $\theta$ as the emissivity in the classical skin effect region ($\lambda < 7\,\mu m$ and $>14\,\mu m$). Fig. 4 shows that effectively the broad peak width is independent of the emission angle, and Fig. 5 shows the experimental emissivity data as a function of the emission angle for 7, 9, and 11 $\mu m$ wavelengths. It can be observed that, in the three plots, the emissivity shows the classical dependence on the angle. The distance between each curve shows a nearly negligible increase with the incidence angle, in agreement with the theoretical predictions.[3,10] The experimental results confirm the very small influence of the incidence angle on the anomalous skin effect, as well as the excellent quality of the experimental measurements.

In this Letter, we have investigated the spectral emissivity in the regime of the anomalous skin effect above room temperature. We have found that the normal spectral emissivity presents a broad peak around 10 $\mu m$. It has been checked that the influence of the emission angle on the broad peak values follows the behaviour predicted by the general electromagnetic theory. The experimental results confirm that the influence of the emission angle is not significant on the anomalous skin effect regime. The experimental results are in excellent qualitative agreement with the broad peak predicted by the anomalous skin effect in the medium infrared with $p = 1$. This effect has not been observed in silver films reflectance measurements where the experimental data are in good agreement with the simple Drude theory predictions.[33] In addition, experimental work is being carried out to study the anomalous skin effect regime in noble metals at low temperatures. Therefore, further theoretical emissivity calculations for intermediate $p$ values as well as an experimental study about the dependence of emissivity on roughness are necessary. Finally, we can say that the highly accurate emissivity measurements together with the theoretical simulations used to obtain optical constants open up an alternative way to study optical constants of metals, and, in particular, the role of the gain term in the collision integral of the Boltzmann equation, which seems to be specially important in the intermediate region $\ell \approx \delta$.[12]

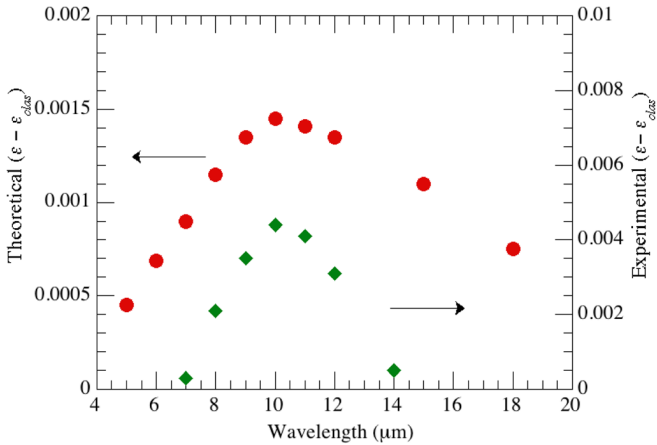

FIG. 3. Comparison between $(\epsilon_{peak} - \epsilon_{clas})$ plots for both experimental results and theoretical predictions. The experimental $\epsilon_{clas}$ values used in this plot were obtained with a base line between 7 and 14 $\mu m$.

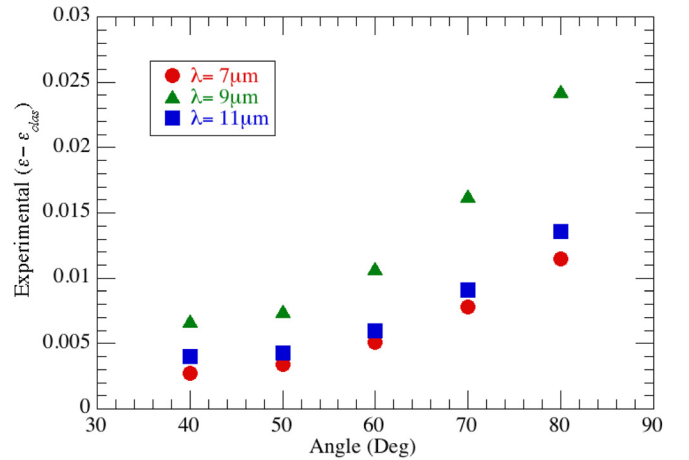

FIG. 5. Emissivity data as a function of the incidence angle for three wavelengths. The three curves show the classical parabolic behaviour. The nearly negligible differences are in agreement with the theoretical predictions of the anomalous skin effect for oblique incidence angles.

This research was partially supported by the program ETORTEK 2011 of the Consejería de Industria of the Gobierno Vasco in collaboration with the CIC-Energigune Research Center.